\documentclass[12pt]{article}
\usepackage{layout,amssymb,amsmath,epsfig}

\begin{document}

\title{\bf Strange Stars in $f(T)$ Gravity With MIT Bag Model}
\author{G. Abbas$^a$ \thanks{%
abbasg91@yahoo.com} \ Shahid Qaisar$^a$ \thanks{%
shahidqaisar90@ciitsahiwal.edu.pk}  and Abdul Jawad$^b$ \thanks{%
sweetfriend181@gmail.com }\\\\
%EndAName
 $^a$Department of
Mathematics, COMSATS \\ Institute
of Information Technology, Sahiwal-57000, Pakistan.\\
$^b$Department of Mathematics, COMSATS
\\Institute of Information
Technology, Lahore, Pakistan.}
\date{}
\maketitle
\begin{abstract}
This paper deals with existence of strange stars in $f\left(
T\right) $ modified gravity. For this purpose, we have taken the
diagonal tetrad field of static spacetime with charged anisotropic
fluid and MIT bag model, which provide the linear relation between
radial pressure and density of the matter. Further, the analysis of
the resulting equations have been done by assuming the parametric
form of the metric functions in term of the radial profiles with
some unknown constant (introduced by Krori and Barua). By the
matching of two metrices, unknown constant of the metric functions
appear in terms of mass, radius and charge of the stars, the
observed values of these quantities have been used for the detail
analysis of the the derived model. We have discuss the regularity,
anisotropy, energy conditions, stability and surface redshift of the
model.
\end{abstract}


\section{Introduction}
Recently, the modified theories of gravitation have attracted the
attention of a huge community of theoretical physicists due to their
implication as an alternative framework for the successful proposal
of dark energy. This modification is carried out in several ways by
modifying the Lagrangian of General Relativity (GR), for examples
Ricci scalar $R$ in the Einstein-Hilbert action must be replaced by
$f\left( R\right)$ (see Durrer and Maartens 2010; DE Felice and
Tsujikawa 20210). In GR and modified theories of gravity, the
gravitational effects could be described in term of contents of
matter and curvature of spacetime. However, one can formulate an
analogous theory of GR, which can be demonstrated in term of torsion
instead of curvature. Originally, Einstein formulated this theory to
unify the basic forces due to Coulomb field and gravitational field
Einstein (1928). This problem of unification remains unsolved in the
new theory, but after many years the theory becomes much popular as
an alternative theory of  GR , this theory is known as teleparallel
equivalent of GR (TEGR) M$\ddot{o}$ller (1961). The basic concept to
formulate this theory is to take a such manifold that preserve
torsion along with curvature. The complete Riemann curvature tensor
can be considered as zero, hence in this way gravitational field can
be described by torsion. The appropriate technique is to utilize the
tetrad fields $e_{\mu }^{i}$ Weitzenb$\ddot{o}$ck (1923).

During the current years, the researchers have carried out the study
of TEGR with some modification to explore some theoretical as well
experimental problems of modern cosmology (Ferraro and Fiorini 2007,
2008). The comparison of $f\left( R\right)$ gravity with $f(T)$
gravity implies that later one is more convenient as in $f(T)$
gravity, equations of motions are of second order while for metric
$f\left( R\right) $ gravity, these equations are fourth order.
Recently, a lot of work in $f\left( T\right)$ gravity is devoted to
investigate the observed data of the theoretical cosmology or to
determine the constraint on the parameter of theory, so that the
theoretical results may fit with the data obtained by some standard
source like WMAP etc. Further, many researchers have investigated
that a particular class of $f\left( T\right) $ gravity models can
explain some basic problems of cosmology like inflation and
accelerated expansion (Bengochea and Ferraro 2009 Wu and Yu 2010).

In addition to study the cosmology, $f(T)$ theory has been used to
examined the effects of $f(T)$ models on the existence of Black hole
and compact objects in 3-dimensions. On the basis of this fact, the
 BTZ black hole model has been derived in $f(T)$ theory of gravity.
(Zheng and Huang 2011). Also, it has been proved that the first of
black hole thermodynamics in $f(T)$ gravity becomes invalid, due to
the break down of Lorentz invariance (Bengochea 2011). Further, some
static vacuum solutions with charged source have been investigated
in $f(T)$ theory, such solutions leads to charged black hole
solutions in $f(T)$ gravity (Li et al.2011). A large class of static
perfect fluid solutions exhibit the existence of relativistic stars
in $f(T)$ gravity ( Dent et al.2011, Rong-Xin et al.2011; Wang et
al. 2011, Boehmer et al. 2011, Daouda 2011).

It has been the subject of great interest to study the models of
anisotropic stars during the last decades  in GR as well in modified
theories of gravity (Herrera and Santos 1997, Chan et al.1993, Di
Prisco 1997, Herrera and Barreto (2013), Abbas (2014a), (2014b),
(2014c), (2014d), Abbas and Sabiullah (2014)). Mak and Harko (2004)
present a class of regular and analytic solution of field equation
with anisotropic source. Also, Chaisi and Maharaj (2005) have
discussed the nature of anisotropic matter source, after
establishing an algorithm. By utilizing the Chaplygin gas equation
of state (EOS), Rahaman et al. (2012) generalized the Krori-Barua
(KB) (1975) solutions to the matter source with electric field. It
has become a scientific tool to study the models the compact stars
with Krori-Barua metric (Kalam et al. 2012, Kalam et al. 2013).
Hossein et al. (2012) discussed the modelling of the compact star
model by taking cosmological constant as radial coordinate
dependent. Bhar et al. (2015) and  Maurya et al.(2015) discuss the
possibility for the existence of higher dimensional and charged
compact stars, respectively.

In present year, Abbas and his collaborators ( Abbas et al.2014,
Abbas et al.2015a, 2015b, 2015c, 2015d ) have studied the models of
anisotropic compact stars in GR, $f(R)$, $f(G)$ and $f(T)$ theories
in diagonal tetrad case using Krori and Barua (KB) metric approach.
Das et al.(2015) have examined the existence of compact stars in
$f(T)$ gravity using the conformal motion technique and have
discussed properties of the resulting models. The main objective of
the present paper is to study the models of the anisotropic compact
stars in the context of $f(T)$ gravity using diagonal tetrad in the
presence of electric field and MIT Bag Model. We have examined the
anisotropic behavior, regularity at the center as well as stability
of these models. Finally, the surface redshift has been calculated.
All these properties of the models have been discussed by using the
observational data of the compact stars. The plan of the paper is
the following:  In the next section, we present the review of $f(T)$
gravity and equations of motion with charged anisotropic. Section
\textbf{3} deals with charged anisotropic source and the equation of
motion in $f(T)$ gravity with diagonal tetrad. Section \textbf{4}
investigates with the physical analysis of the proposed models. In
the last section, we conclude the results of the paper.

\section{$f(T)$ Theory of Gravity}

Currently, Teleparallel theory of gravity is an equivalent theory of
gravitation to the GR (Zhang et al.2011, Li et al. 2011). In this
section, we shall introduce the basic concepts of the $f(T )$
gravity theory. To this end, we define notation of the Latin indices
for the tetrad field and Greek indices for spacetime coordinates.
The line element of the manifold is given by
\begin{equation}
ds^{2}=g_{\mu\nu}dx^{\mu}dx^{\nu}.
\end{equation}
This spacetime can be transformed to a Minkowskian form by the
matrix transformation as follows:
\begin{equation}
dS^{2}=g_{\mu\nu}dx^{\mu}dx^{\nu}=\eta_{ij}\theta^{i}\theta^{j},
\end{equation}
\begin{equation}
dx^{\mu}=e_{i}^{\mu}\theta^{i}, \theta^{i}=e^{i}_{{\mu}}dx^{\mu},
\end{equation}
where $\eta_{ij}=diag[1,-1,-1,-1]$ and $e_{i}^{\mu}e_{i}^{\nu}=\delta_{%
\nu}^{\mu}$ or $e_{i}^{\mu}e_{j}^{\nu}=\delta_{i}^{j}$. \newline
The root of the metric determinant is given by $\sqrt{-g }%
=det[e_{\mu}^{i}]=e. $

We consider the manifold for which the Riemann tensor is zero and
non-zero torsion terms exist, the Weitzenbock's connection
components can be defined as follows:
\begin{equation}
\Gamma^{\alpha}_{\mu\nu}=e_{i}^{\alpha}\partial_{\nu}e_{\mu}^{i}.
\end{equation}
We define the torsion and the contorsion tensor as follows:\newline
\begin{equation}
T^{\alpha}_{\mu\nu}=\Gamma^{\alpha}_{\nu\mu}-\Gamma^{\alpha}_{\mu%
\nu}=e_{i}^{\alpha}(\partial_{\mu}e_{\nu}^{i}-\partial_{\nu}e_{\mu}^{i}),
\end{equation}
\begin{equation}
K^{\mu\nu}_{\alpha}=-\frac{1}{2}(T^{\mu\nu}_{\alpha}-T^{\nu\mu}_{\alpha}-T^{%
\mu\nu}_{\alpha}),
\end{equation}
and the components tensor $S_{\alpha}^{\mu\nu}$ are defined as
\begin{equation}
S_{\alpha}^{\mu\nu}=\frac{1}{2}(K^{\mu\nu}_{\alpha}+\delta^{\mu}_{\alpha}T^{%
\beta\nu}_{\beta}-\delta^{\nu}_{\alpha}T^{\beta\mu}_{\beta}),
\end{equation}
one can write the torsion scalar as\newline
\begin{equation}
T=T^{\alpha}_{\mu\nu}S_{\alpha}^{\mu\nu}.
\end{equation}
Now, similarly to the $f(R)$ gravity, one defines the action of
$f(T)$ theory as follows:\newline
\begin{equation}
S[e_{\mu}^{i}, \Phi_A]=\int d^4 x e[\frac{1}{16\pi}f(T)+\mathcal{L}%
_Matter(\Phi_A)],
\end{equation}
where we used $G=c=1$ and $\Phi_A$ is matter fields. Now the
variation of above action yields the following form of equations of
motion (Wu and Yu  2010 and Zheng and Huang 2011)
\begin{equation}\label{Em10}
S_{\mu}^{\nu\rho}\partial_{\rho}T
f_{TT}+[e^{-1}e^{i}_{\mu}\partial_{\rho}(ee_i^{\alpha}S_{\alpha}^{\nu%
\rho})+T^{\alpha}_{\lambda\mu} S_{\alpha}^{\nu\lambda}]f_T+\frac{1}{4}%
\delta^{\nu}_{\mu}f=4\pi\left(
\mathcal{T}_{\mu}^{\nu}+E_{\mu}^{\nu}\right),
\end{equation}
where $\mathcal{T}_{\mu}^{\nu}$ and $E_{\mu}^{\nu}$ are energy
momentum tensors of ordinary matter and electromagnetic field.

The ordinary matter is an anisotropic fluid for which the
energy-momentum tensor is given by\newline
\begin{equation}
\mathcal{T}_{\mu }^{\nu }=(\rho +p_{t})u_{\mu }u^{\nu }-p_{t}\delta _{\mu
}^{\nu }+(p_{r}-p_{t})v_{\mu }v^{\nu },
\end{equation}%
where $u^{\mu }$ is the four-velocity, $v^{\mu }$ radial four
vector, ${\rho }$ the energy density, ${p_{r}}$ is the radial
pressure and ${p_{t}}$ is transverse pressure. Further, the energy
momentum tensor for electromagnetic field is given by

\begin{equation}\label{1.3.8}
E_{{\mu}^{\nu}}=\frac{1}{4{\pi}}(g^{{\delta}{\omega}}
F_{{\mu}{\delta}}F^{\nu}{\omega}-\frac{1}{4}g_{\mu}^{\nu}
F_{{\delta}{\omega}}F^{{\delta}{\omega}}),
\end{equation}
where $F_{{\mu}{\nu}}$ is the Maxwell field tensor defined as
\begin{equation}\label{1.3.9}
F_{{\mu}{\nu}}={\phi}_{\nu, \mu}-{\phi}_{\mu, \nu}
\end{equation}
and ${\phi}_{\mu}$ is the four potential.

\section{Equation of Motion in $f(T)$ Gravity }

We assume that the interior spacetime of a strange star is describe
by the KB (1975) metric

\begin{equation}\label{1}
ds^{2}=-e^{a(r)}dt^{2}+e^{b(r)}dr^{2}+r^{2}(d{\theta ^{2}}+sin^{2}d\phi ^{2})
\end{equation}

\bigskip In order to re-write above line element, into the invariant form
under the Lorentz transformations, we define the tetrad matrix as
\begin{equation}
\lbrack e_{\mu }^{i}]=diag[e^{\frac{a(r)}{2}},e^{\frac{b(r)}{2}%
},r,rsin(\theta )].
\end{equation}
Further, one can obtain $e=det[e_{\mu }^{i}]=e^{\frac{(a+b)}{2}%
}r^{2}sin(\theta )$. For the charged fluid source with density $\rho
(r)$ radial pressure $p_{r}(r)$, tangential pressure $p_{t}(r)$. The
Einstein-Maxwell (EM) equations take the form ( with geometrized
units $G=c=1$).

\begin{eqnarray}
T(r) &=&\frac{2e^{-b}}{r}(a^{^{\prime }}+\frac{1}{r}), \\
T^{^{\prime }}(r) &=&\frac{2e^{-b}}{r}(a^{^{\prime \prime }}+\frac{1}{r^{2}}%
-T(b^{^{\prime }}+\frac{1}{r})),
\end{eqnarray}
where the prime $^{\prime }$ denotes the derivative with respect to
the radial coordinate $r$. One can now re-write the equations of
motion for anisotropic fluid as
\begin{eqnarray}\label{EM1}
4\pi \rho +E^{2}&&=\frac{f}{4}-\left( T-\frac{1}{r^{2}}-\frac{e^{-b}}{r}%
(a^{^{\prime }}+b^{^{\prime }})\right) \frac{f_{T}}{2},\\\label{EM2}
 4\pi
p_{r}-E^{2}&&=\left( T-\frac{1}{r^{2}}\right)
\frac{f_{T}}{2}-\frac{f}{4},\\\label{EM3}
4\pi p_{t}+E^{2}&&=\left[ \frac{T}{2}+e^{-b}\left( \frac{a^{^{\prime \prime }}%
}{2}+(\frac{a^{^{\prime }}}{4}+\frac{1}{2r})(a^{^{\prime
}}-b^{^{\prime }})\right) \right]
\frac{f_{T}}{2}-\frac{f}{4},\\\label{EM4} \frac{cot\theta
}{2r^{2}}T^{^{\prime }}f_{TT}&&=0,\\\label{EM5}
E(r)&&=\frac{1}{r}\int_{0}^{r}4{\pi }r^{2}\sigma {e^{\frac{\lambda }{2}}}dr=%
\frac{q(r)}{r^{2}}
\end{eqnarray}

where $q\left( r\right) $ is the total charge within a sphere of
radius $r$. According to the MIT bag model, we take the simple form
of the strange matter EoS (Rahman et al.(2012))
\begin{equation}
p_{r}=\frac{1}{3}(\rho -4B_{g}).
\end{equation}

From Eq.(\ref{EM4}), we get the following linear form of $f(T)$
function
\begin{equation}
f(T)=\beta T+{\beta }_{1},
\end{equation}%
where $\beta $ and ${\beta }_{1}$ are integration constants, for
simplicity we assume ${\beta }_{1}=0$. We parameterize the metric as
follows (Krori and Barua 1975)
\begin{equation}\label{EM6}
b(r)=Ar^{2},\quad a(r)=Br^{2}+C,
\end{equation}%
where $A$, $B$ and $C$ are are some arbitrary constants which will
be determined later using some physical conditions. With the choice
of the EoS, we have a
system of five independent equations with five unknown parameters namely, $%
\rho $, $p_{r},p_{t},E(r)$ and $\sigma (r)$. From
Eqs.(\ref{EM1})-(\ref{EM6}), we obtain

\begin{eqnarray}\label{EMN1}
{\rho }&=&\frac{3(A+B)}{16\pi }e^{{-}Ar^{2}}\beta
+B_{g},\\\label{EMN2} p_{r}&=&\frac{(A+B)}{16\pi }e^{-Ar^{2}}\beta
-B_{g},\\\nonumber
p_{t} &=&\frac{1}{8\pi }\left[\left(\frac{7}{2}B-\frac{3}{2}A+B^{2}r^{2}-ABr^{2}+\frac{%
1}{r^{2}}\right)e^{-Ar^{2}}\beta -\frac{1}{r^{2}}\right]
\\\label{EMN3} &&+ B_{g},\\\label{EMN4}
E^{2}&=&\frac{1}{2}(A-3B-\frac{2}{r^{2}})e^{-Ar^{2}}\beta +\frac{\beta }{r^{2}}%
-8{\pi }B_{g},
\end{eqnarray}
The charge density is obtained as
\begin{equation}
\sigma =\frac{e^{\frac{-Ar^{2}}{2}}}{2{\pi }r}\psi + \frac{Ae^{\frac{%
-3Ar^{2}}{2}}}{8{\pi }r\psi }\left[ 2-\left( A-3B\right) r^{2}\right]
+ \frac{e^{\frac{-Ar^{2}}{2}}}{4{\pi }r^{3}\psi }\left( e^{-Ar^{2}}-1%
\right) ,
\end{equation}%
where
\begin{equation}
\psi =\sqrt{\left[ \frac{1}{2}(A-3B-\frac{2}{r^{2}})e^{-Ar^{2}}\beta +%
\frac{\beta }{r^{2}}-8{\pi }B_{g}\right] }
\end{equation}%
The amount of net charge inside a sphere of radius $r$ is given by
\begin{equation}
q=r^{2}\sqrt{\left[ \frac{1}{2}(A-3B-\frac{2}{r^{2}})e^{-Ar^{2}}\beta +%
\frac{\beta }{r^{2}}-8{\pi }B_{g}\right] }
\end{equation}%
\section{Physical Analysis of Model}

\subsection{Central Regularity of Model Parameters}

It well know fact that the KB metric preserve no singularity in its
metric functions even at the center at $r=0$. Therefore, it becomes
necessary to impose the constraints on the constants appearing in
the metric functions, so that physical parameters of the model
remain well behaved at all the inner points strange stars. For the
regularity at the center $(r=0)$, we obtain the central density in
the form
\begin{equation}
\rho _{0}=\rho (r=0)=\frac{3(A+B)}{16{\pi }}\beta +B_{g}.
\end{equation}%
Further, the regularity of electric field requires that it must
vanish at the center i.e.,
\begin{equation}
E^{2}(r=0)=\frac{3}{2}(A-B)-8{\pi }B_{g}=0,
\end{equation}%
which implies that
\begin{equation}
B_{g}=\frac{3(A-B)}{16\pi }\beta
\end{equation}
The pressures and density should be decreasing function of $r.$ In
our model, radial variation of $p_{r}$ is obtained as
\begin{equation}
\frac{dp_{r}}{d_{r}}=-\frac{(A+B)}{8\pi }rAe^{-{Ar^{2}}}\beta<0,
\end{equation}%
At $r=0$, $\frac{dp_{r}}{d_{r}}=0$ and $\frac{d^{2}p_{r}}{dr^{2}}%
<0$. This implies that $p_{r}$ will decrease radially outward. The
radial variation of matter energy density is obtained as
\begin{equation}
\frac{d{\rho }}{d_{r}}=-\frac{3(A+B)}{8\pi }rAe^{-{Ar^{2}}}\beta
\end{equation}%
which also shows that at $r=0$,$\frac{d\rho}{d_{r}}=0$ and
$\frac{d^{2}\rho}{dr^{2}}=-\frac{3}{8\pi}\beta(A+B)<0$. Hence
density is decreasing function of $r$, provided that $\beta>0$, as
we have chosen it in our discussion. Hence, it has been shown
graphically that pressures and energy density decreases as shown in
figures \textbf{1-3}.
The anisotropic stress is obtained as%
\begin{eqnarray}\nonumber
\Delta &=&p_{t}-p_{r}=2B_{g}-\frac{\beta }{8{\pi }r^{2}}\\
&&+\frac{1}{8{\pi }}(3B-2A+B^{2}r^{2}-ABr^{2}+\frac{1}{r^{2}}%
)e^{-Ar^{2}}\beta.
\end{eqnarray}

The anisotropic force of the gravitating system will be will be
outwardly directed when $p_{t}>p_{r}$ i.e., $\Delta >0$, while
inwardly directed when $p_{t}<p_{r}$ i.e., $\Delta <0$. It is clear
from the figure \textbf{4} that there exist repulsive gravitational
force for $1\leq\beta>\leq2$ as $(\Delta >0)$, while for
$\beta\geq3$ force becomes attractive. In the case of repulsive
force more massive distribution of matter would exist, which is the
result of disturbance of equilibrium. Figure \textbf{5} implies that
$E^2$ is increasing function of radial coordinate for all the values
of $\beta$.

\begin{figure}
\center\epsfig{file=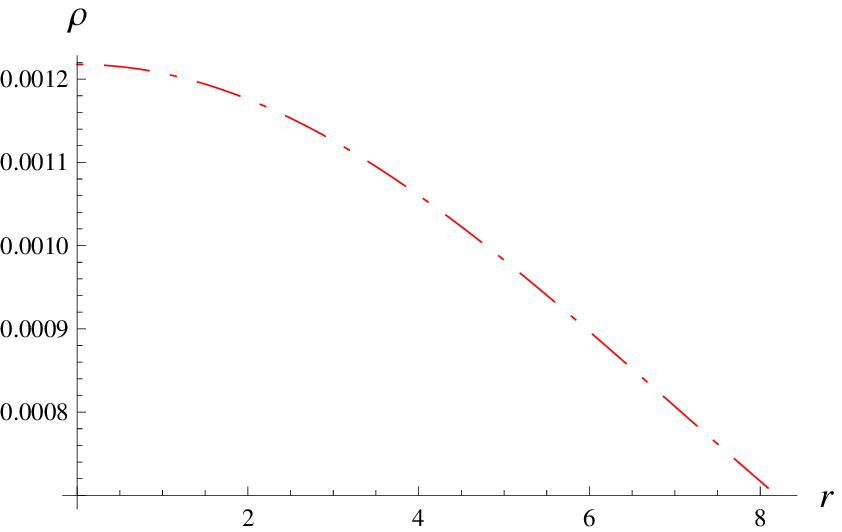, width=0.3\linewidth}
\epsfig{file=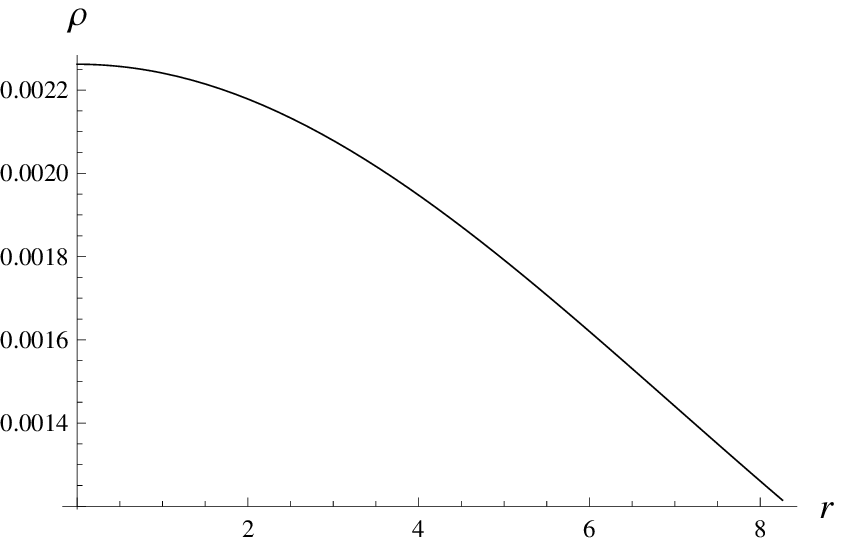, width=0.3\linewidth} \epsfig{file=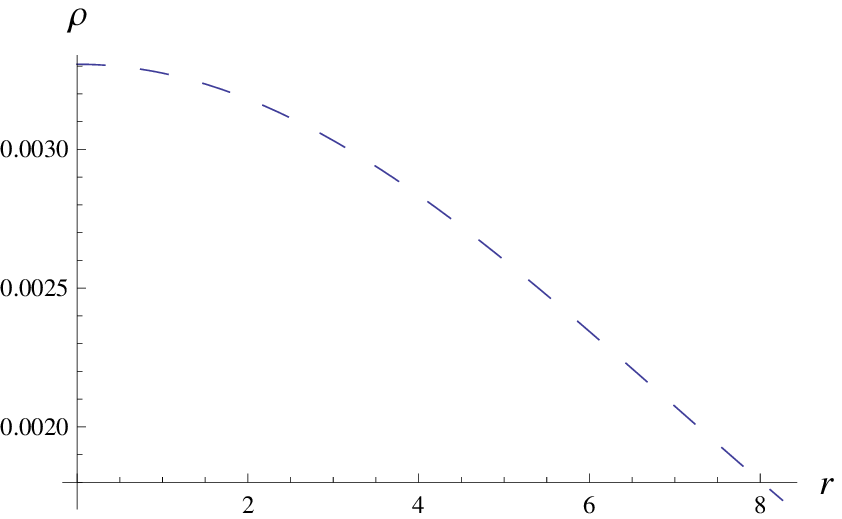,
width=0.3\linewidth} \caption{This figure represents the variation
of $\rho$ versus $r(km)$ for a strange star of radius 8.26 km with
$\beta=1$, $\beta=2$ and $\beta=3$. The corresponding values of $A$,
$B$ and $B_g$ have been used from table \textbf{2}. All the graphs
have been plotted for table \textbf{2}, it will not be mentioned
again.}
\end{figure}
\begin{figure}
\center\epsfig{file=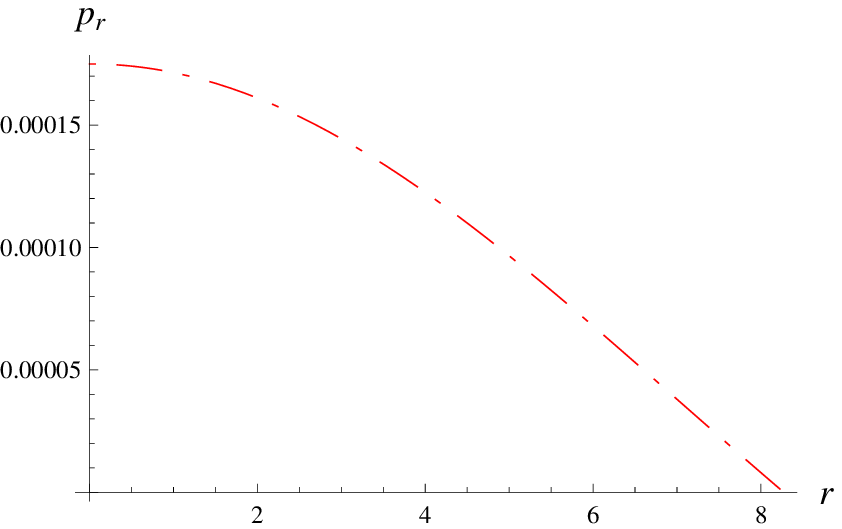, width=0.3\linewidth}
\epsfig{file=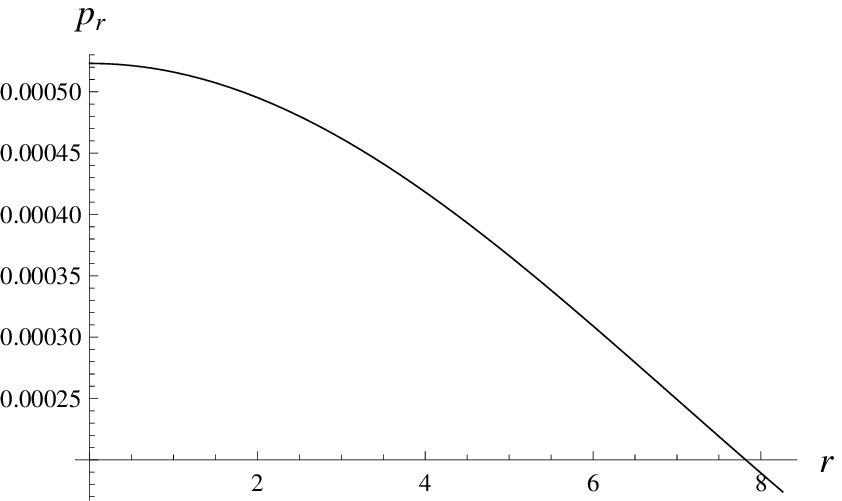, width=0.3\linewidth} \epsfig{file=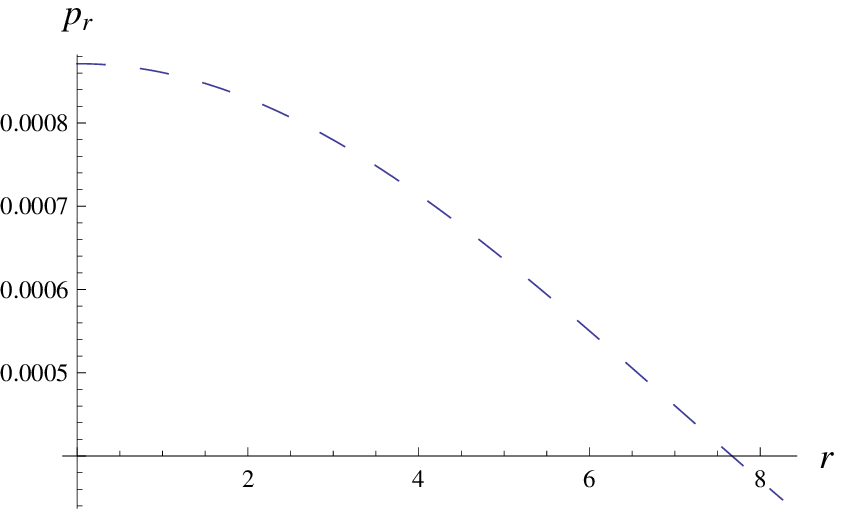,
width=0.3\linewidth} \caption{This figure represents the variation
of $p_r$ versus $r(km)$ for a strange star of radius 8.26 km with
$\beta=1$, $\beta=2$ and $\beta=3$.}
\end{figure}

\begin{figure}
\center\epsfig{file=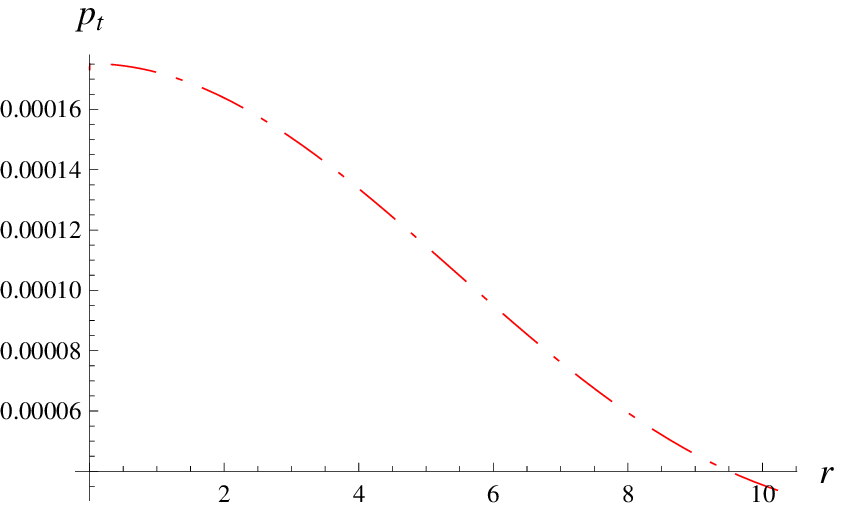, width=0.3\linewidth}
\epsfig{file=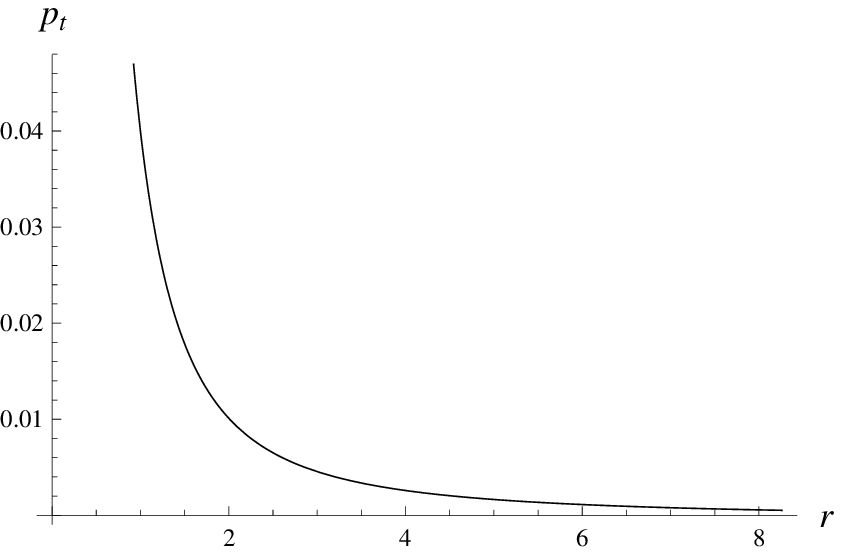, width=0.3\linewidth} \epsfig{file=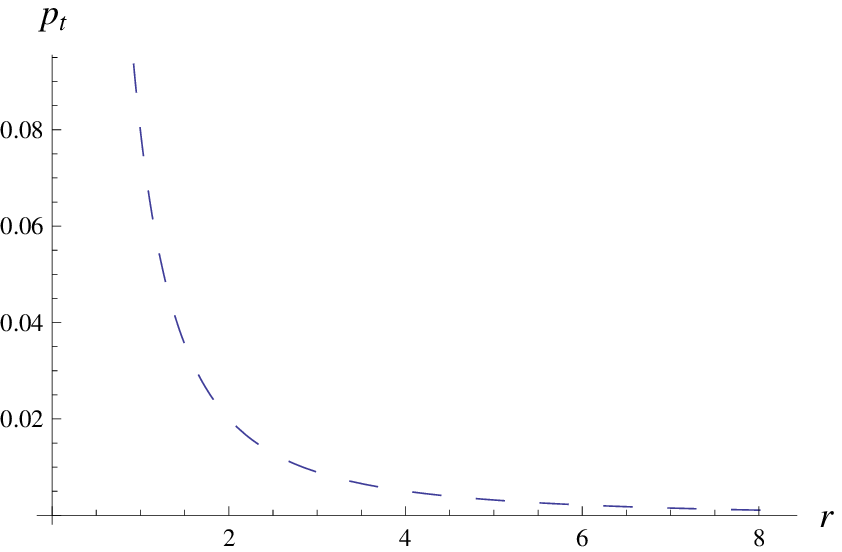,
width=0.3\linewidth} \caption{This figure represents the variation
of $p_t$ versus $r(km)$ for a strange star of radius 8.26 km with
$\beta=1$, $\beta=2$ and $\beta=3$.}
\end{figure}

\subsection{Matching Conditions}

The exterior region of the star is taken as Reissner-Nordstrom
metric given by
\begin{equation}
ds^{2}=-\left( 1-\frac{2M}{r}+\frac{Q^{2}}{r^{2}}\right) dt^{2}+\left( 1-%
\frac{2M}{r}+\frac{Q^{2}}{r^{2}}\right) ^{-1}dt^{2}+r^{2}(d{\theta }%
^{2}+sin^{2}{\theta }d\phi ^{2})
\end{equation}%
where $Q$ is the total charge within boundary $r=R$. Continuity of
the metric coefficients $g_{t}t,$ $g_{r}r$ and $\frac{{\partial }g_{tt}}{{%
\partial }r}$ across the boundary surface $r=R$ between the interior and the
exterior regions of the star yields the following results:
\begin{equation}
1-\frac{2M}{R}+\frac{Q^{2}}{R^{2}}=e^{BR^{2}+C}
\end{equation}%
\begin{equation}
1-\frac{2M}{R}+\frac{Q^{2}}{R^{2}}=e^{-AR^{2}}
\end{equation}%
\begin{equation}
\frac{M}{R^{2}}-\frac{Q^{2}}{R^{3}}=BRe^{BR^{2}+C}
\end{equation}%
 By solving the above three equations, we get
\begin{equation}
A=-\frac{1}{R^{2}}ln\left[ 1-\frac{2M}{R}+\frac{Q^{2}}{R^{2}}\right]
\end{equation}%
\begin{equation}
B=\frac{1}{R^{2}}\left[ \frac{M}{R^{2}}-\frac{Q^{2}}{R^{2}}\right] \left[ 1-%
\frac{2M}{r}+\frac{Q^{2}}{r^{2}}\right] ^{-1}
\end{equation}%
\begin{equation}
C=ln\left[ 1-\frac{2M}{R}+\frac{Q^{2}}{R^{2}}\right] -\frac{\frac{M}{R}-%
\frac{Q^{2}}{R^{2}}}{\left[ 1-\frac{2M}{R}+\frac{Q^{2}}{R^{2}}\right] }
\end{equation}%
For the given values of the parameters $M, R$ and $Q$ the values of
$A$ and $B$ is are given in table \textbf{2}.

\begin{table}[th]
\begin{center}
\begin{tabular}{|c|c|c|c|c|}
\hline {Case} & \textbf{\ $M$} & \textbf{$R(km)$} & \textbf{\
$\frac{M}{R}$} & \textbf{\ }$\frac{Q^{2}}{R^{2}}$ \\ \hline 1 &
1.4$M_{\odot }$ & 8.26 & 0.25 & 0.004 \\ \hline 2 & 1.4$M_{\odot }$
& 6.88 & 0.30 & 0.027 \\ \hline 3 & 1.4$M_{\odot }$ & 5.90 & 0.35 &
0.061 \\ \hline 4 & 1.4$M_{\odot }$ & 5.16 & 0.40 & 0.105 \\ \hline
\end{tabular}%
\end{center}
\caption{Values of $M, R$ and $Q$ defined in Rahman et al.(2012)}
\end{table}
\begin{table}[th]
\begin{center}
\begin{tabular}{|c|c|c|c|}
\hline {Case} & \textbf{\ $A$ $\left( km^{-2}\right) $} & $B$
\textbf{$(km^{-2})$} & \textbf{\ }$\mathbf{B}_{g}$ $(km^{-2})$ \\
\hline 1 & 0.0102 & 0.0073 & 0.0001732 \\ \hline 2 & 0.01798 &
0.01351 & 0.000267 \\ \hline 3 & 0.0292 & 0.0231 & 0.0003643 \\
\hline 4 & 0.044 & 0.037 & 0.000418 \\ \hline
\end{tabular}%
\end{center}
\caption{Values of the model parameters $A,B$ and $B_g$ for given
Masses and Radii of Stars}
\end{table}

\begin{table}[th]
\begin{center}
\begin{tabular}{|c|c|c|c|}
\hline
{Case} & \textbf{\ }$\rho \left( r=0\right) $\textbf{\ $\left( gm\text{ }%
cm^{-3}\right) $} & $\rho \left( r=R\right) $ $\left( gm\text{ }%
cm^{-3}\right) $ & \textbf{\ }$\mathbf{p}_{r}\left( dyne\text{ }%
cm^{-2}\right) $ \\ \hline
1 & 1.643$\times 10^{15}$ & 9.362$\times 10^{14}$ & 2.123$\times 10^{35}$ \\
\hline
2 & 2.895$\times 10^{15}$ & 1.443$\times 10^{15}$ & 4.361$\times 10^{35}$ \\
\hline
3 & 4.703$\times 10^{15}$ & 2.015$\times 10^{15}$ & 8.204$\times 10^{35}$ \\
\hline
4 & 7.087$\times 10^{15}$ & 2.58$\times 10^{15}$ & 14.48$\times 10^{35}$ \\
\hline
\end{tabular}%
\end{center}
\caption{Values of density and pressure at center and surface of
star for $\beta=1$}
\end{table}
\begin{table}[th]
\begin{center}
\begin{tabular}{|c|c|c|c|}
\hline
{Case} & \textbf{\ }$\rho \left( r=0\right) $\textbf{\ $\left( gm\text{ }%
cm^{-3}\right) $} & $\rho \left( r=R\right) $ $\left( gm\text{ }%
cm^{-3}\right) $ & \textbf{\ }$\mathbf{p}_{r}\left( dyne\text{ }%
cm^{-2}\right) $ \\ \hline
1 & 1.643$\times 10^{15}$ & 9.362$\times 10^{14}$ & 2.123$\times 10^{35}$ \\
\hline
2 & 2.895$\times 10^{15}$ & 1.443$\times 10^{15}$ & 4.361$\times 10^{35}$ \\
\hline
3 & 4.703$\times 10^{15}$ & 2.015$\times 10^{15}$ & 8.204$\times 10^{35}$ \\
\hline
4 & 7.087$\times 10^{15}$ & 2.58$\times 10^{15}$ & 14.48$\times 10^{35}$ \\
\hline
\end{tabular}
\end{center}
\caption{Values of density and pressure at center and surface of
star for $\beta=2$}
\end{table}
\begin{table}[th]
\begin{center}
\begin{tabular}{|c|c|c|c|}
\hline
{Case} & \textbf{\ }$\rho \left( r=0\right) $\textbf{\ $\left( gm\text{ }%
cm^{-3}\right) $} & $\rho \left( r=R\right) $ $\left( gm\text{ }%
cm^{-3}\right) $ & \textbf{\ }$\mathbf{p}_{r}\left( dyne\text{ }%
cm^{-2}\right) $ \\ \hline
1 & 1.643$\times 10^{15}$ & 9.362$\times 10^{14}$ & 2.123$\times 10^{35}$ \\
\hline
2 & 2.895$\times 10^{15}$ & 1.443$\times 10^{15}$ & 4.361$\times 10^{35}$ \\
\hline
3 & 4.703$\times 10^{15}$ & 2.015$\times 10^{15}$ & 8.204$\times 10^{35}$ \\
\hline
4 & 7.087$\times 10^{15}$ & 2.58$\times 10^{15}$ & 14.48$\times 10^{35}$ \\
\hline
\end{tabular}%
\end{center}
\caption{Values of density and pressure at center and surface of
star for $\beta=3$}
\end{table}

\begin{figure} \center\epsfig{file=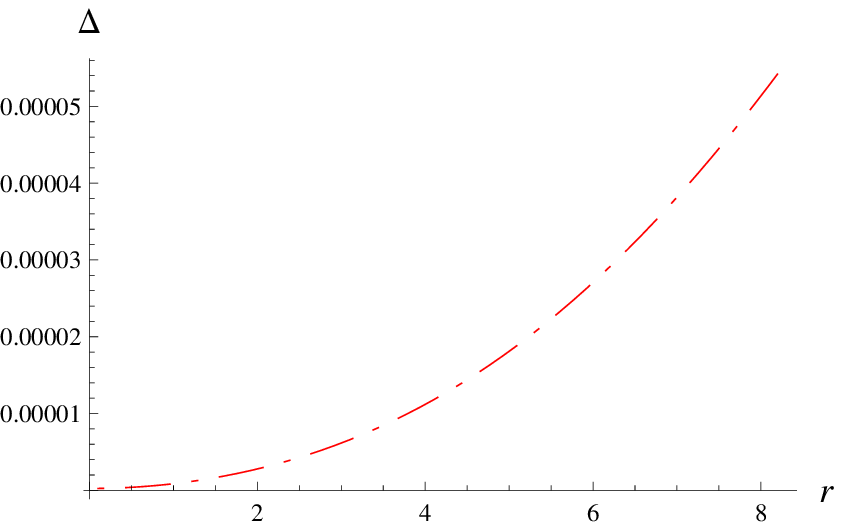, width=0.3\linewidth}
\epsfig{file=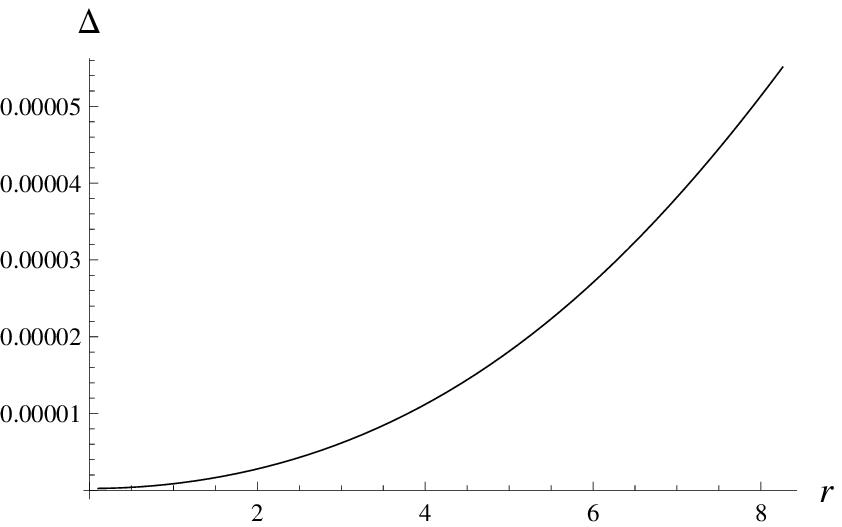, width=0.3\linewidth} \epsfig{file=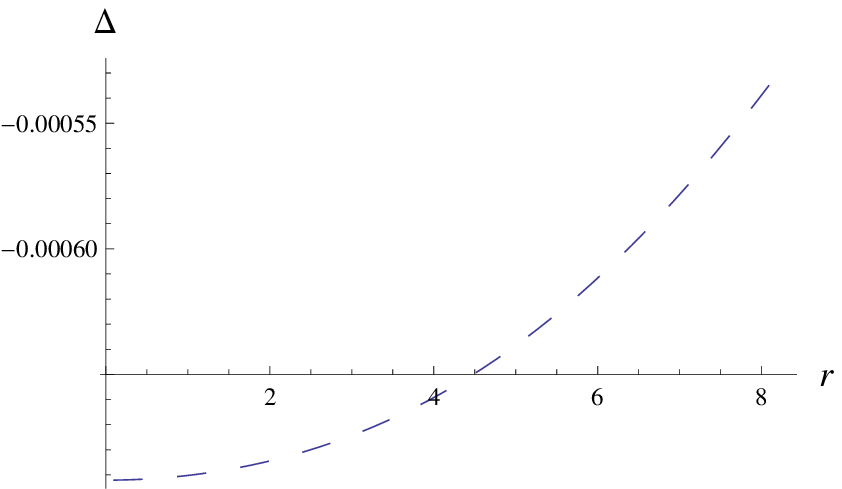,
width=0.3\linewidth}\caption{This figure represents the variation of
$\Delta$ versus $r(km)$ for a strange star of radius 8.26 km with
$\beta=1$, $\beta=2$ and $\beta=3$.}
\end{figure}
\begin{figure}
\center\epsfig{file=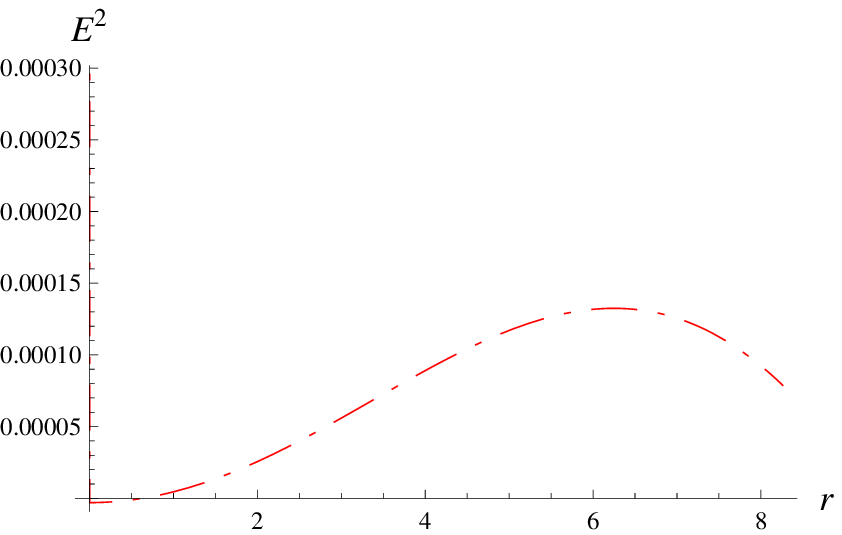, width=0.3\linewidth}
\epsfig{file=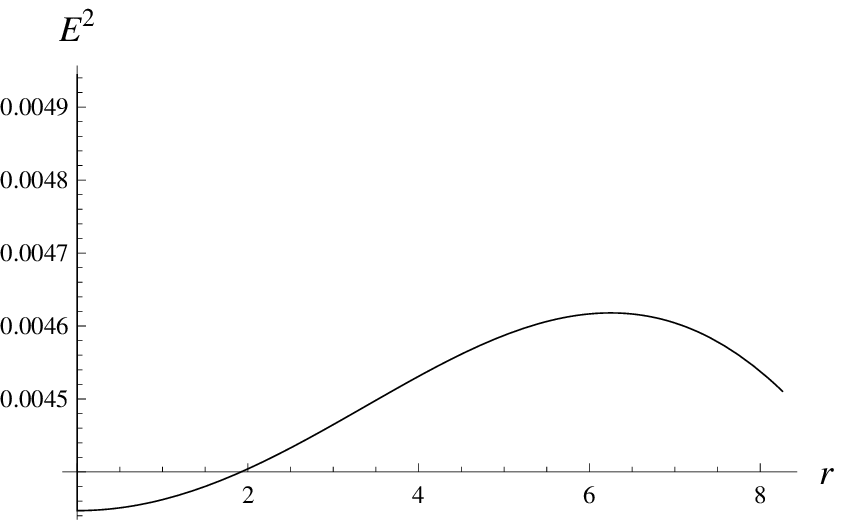, width=0.3\linewidth} \epsfig{file=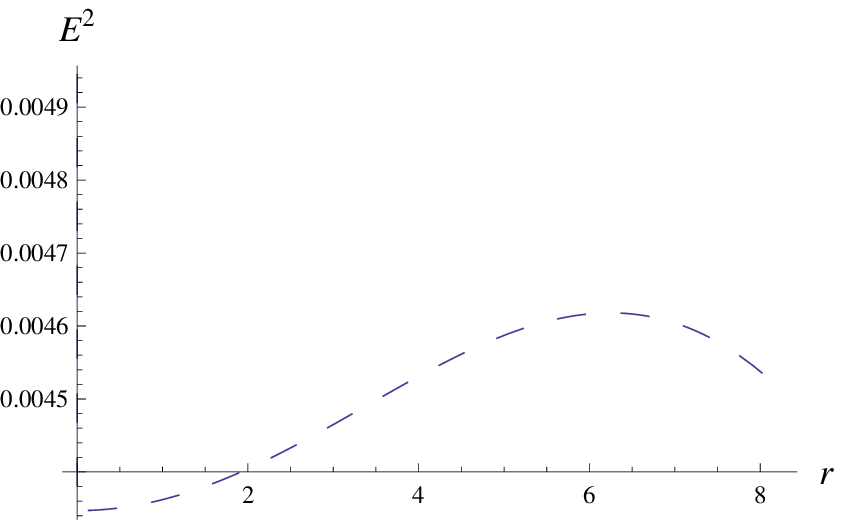,
width=0.3\linewidth}\caption{This figure represents the variation of
${E^2}$ versus $r(km)$ for a strange star of radius 8.26 km with
$\beta=1$, $\beta=2$ and $\beta=3$.}
\end{figure}

\begin{figure}
\center\epsfig{file=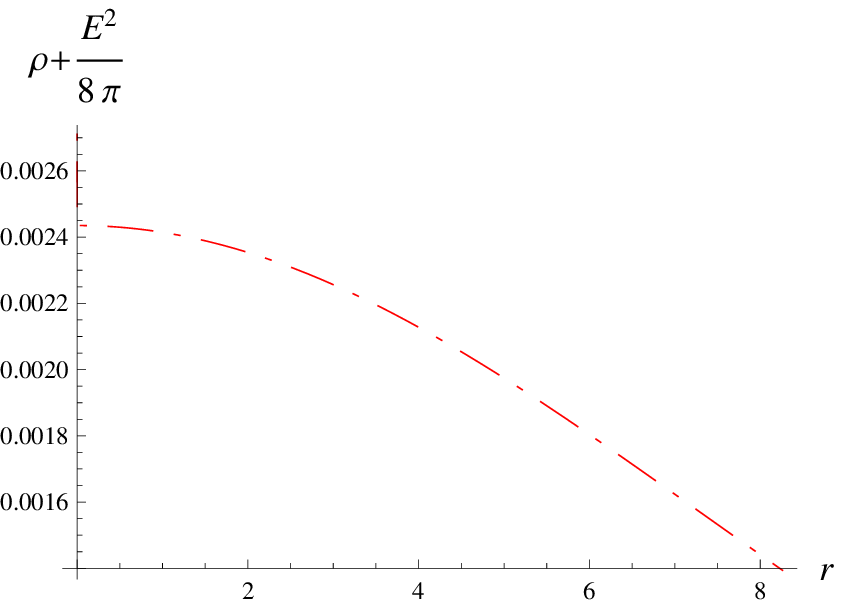, width=0.3\linewidth}
\epsfig{file=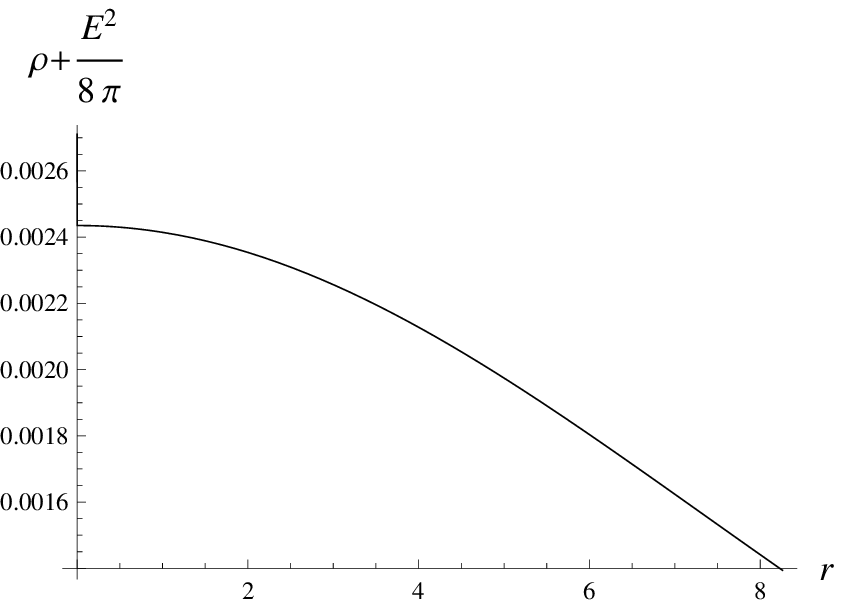, width=0.3\linewidth} \epsfig{file=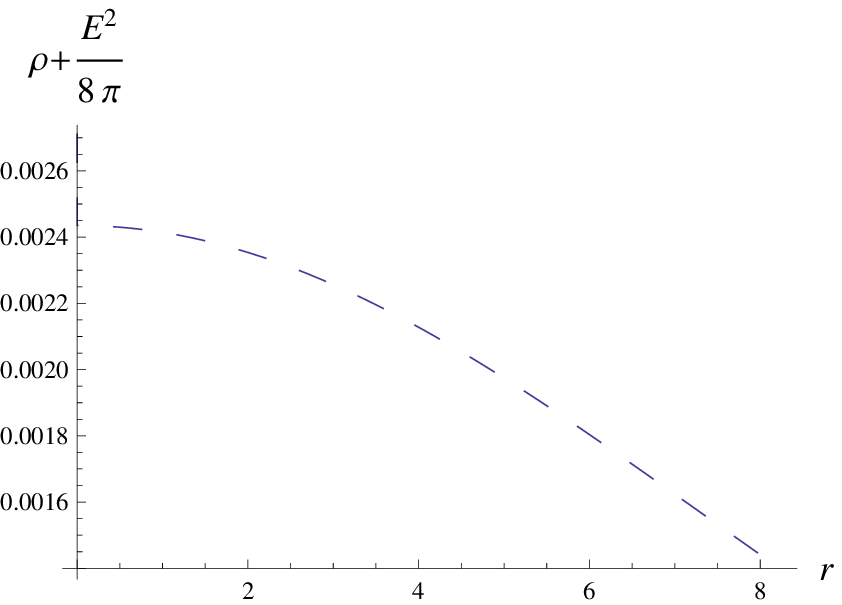,
width=0.3\linewidth} \caption{This figure represents the variation
of $\rho+\frac{E^2}{4\pi}$ versus $r(km)$ for a strange star of
radius 8.26 km with $\beta=1$, $\beta=2$ and $\beta=3$.}
\end{figure}
\begin{figure}
\center\epsfig{file=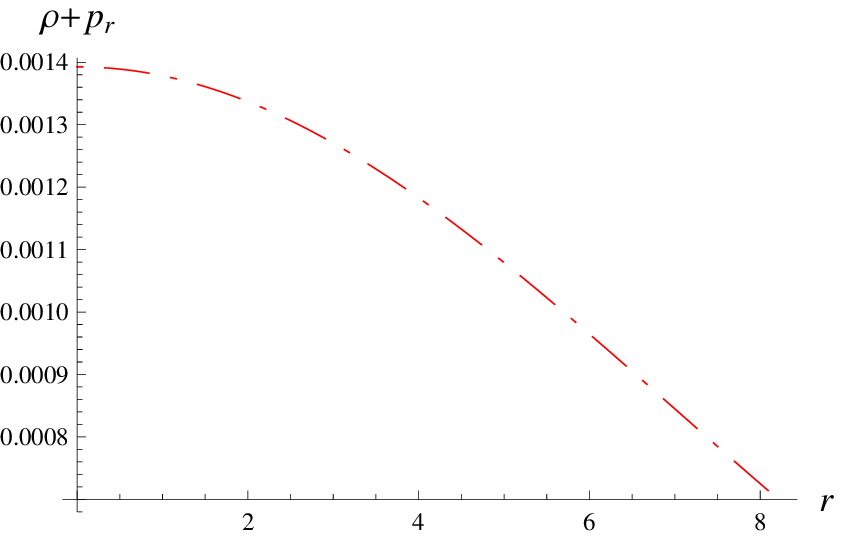, width=0.3\linewidth}
\epsfig{file=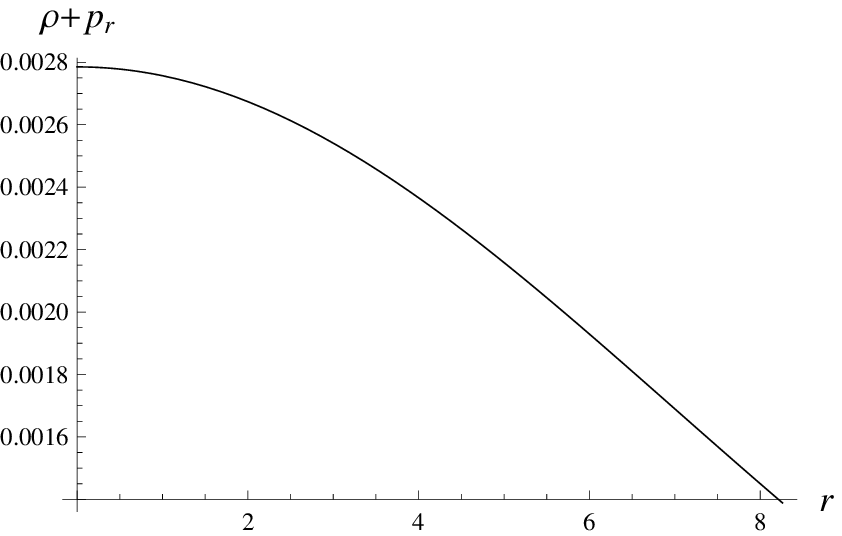, width=0.3\linewidth} \epsfig{file=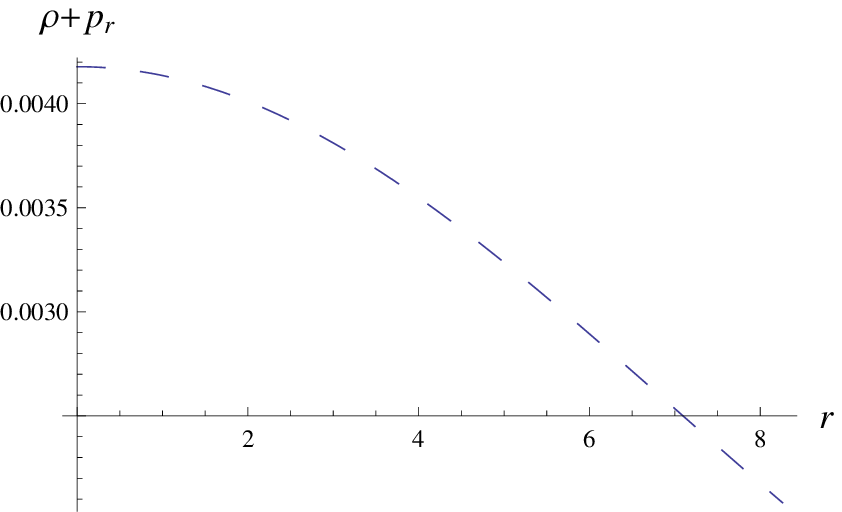,
width=0.3\linewidth} \caption{This figure represents the variation
of $\rho+p_r$ versus $r(km)$ for a strange star of radius 8.26 km
with $\beta=1$, $\beta=2$ and $\beta=3$.}
\end{figure}

\begin{figure}
\center\epsfig{file=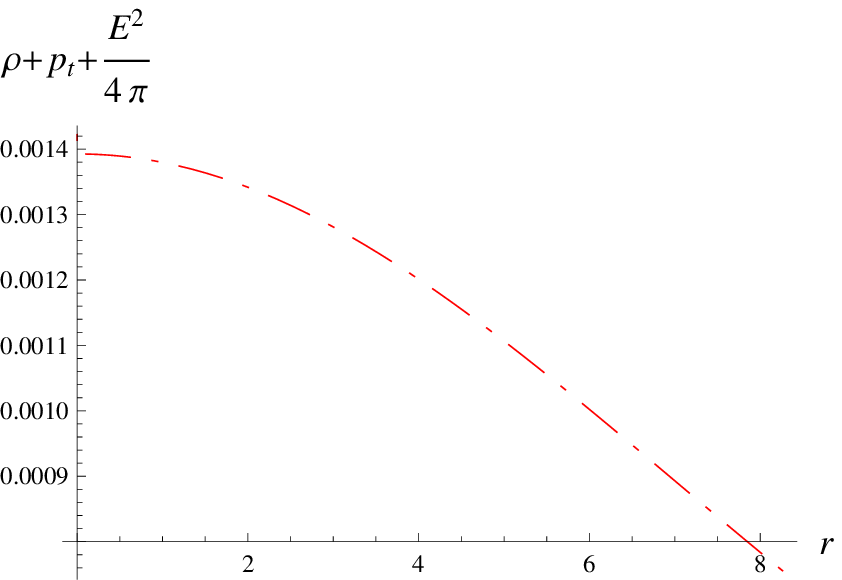, width=0.3\linewidth}
\epsfig{file=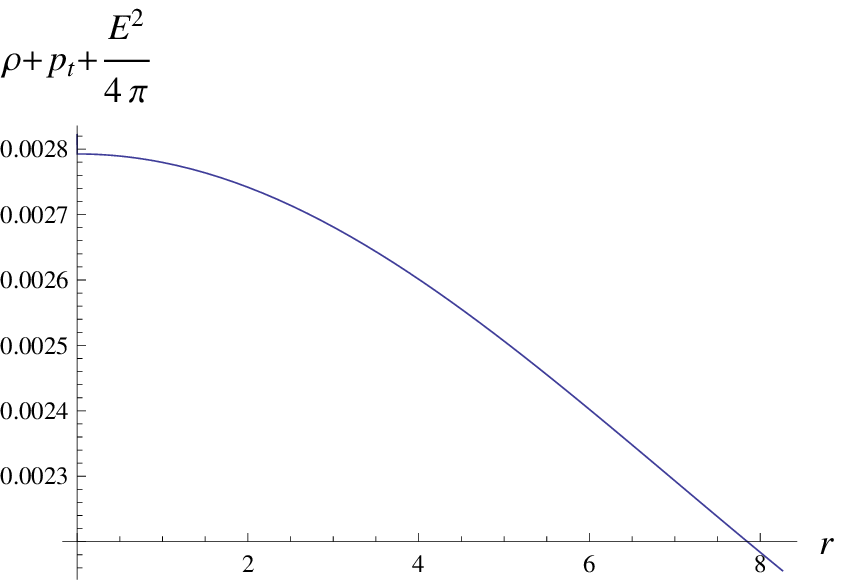, width=0.3\linewidth} \epsfig{file=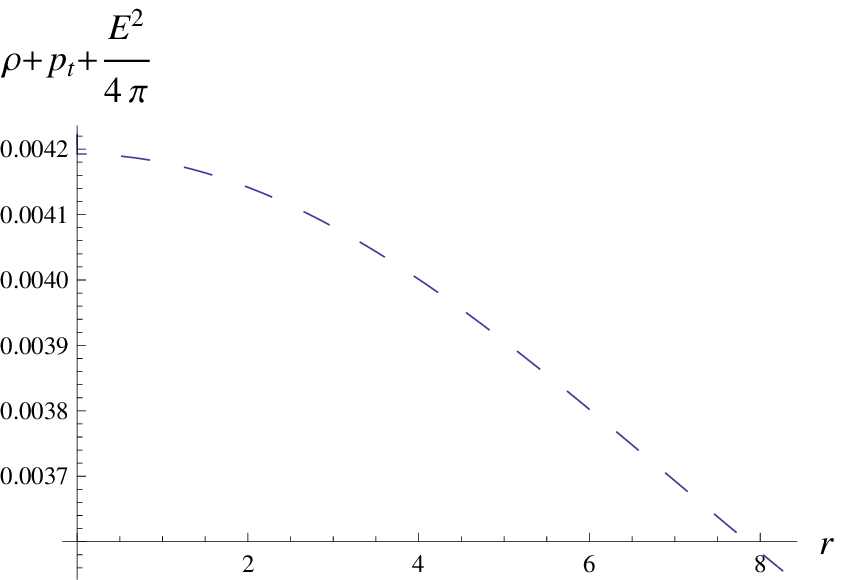,
width=0.3\linewidth} \caption{This figure represents the variation
of $\rho+p_r+\frac{E^2}{4\pi}$ versus $r(km)$ for a strange star of
radius 8.26 km with $\beta=1$, $\beta=2$ and $\beta=3$.}
\end{figure}

\begin{figure}
\center\epsfig{file=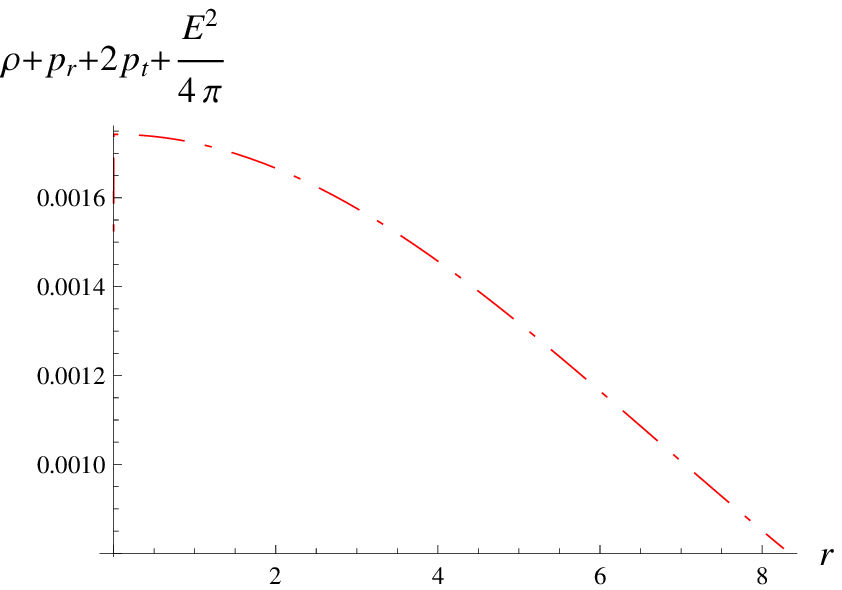, width=0.3\linewidth}
\epsfig{file=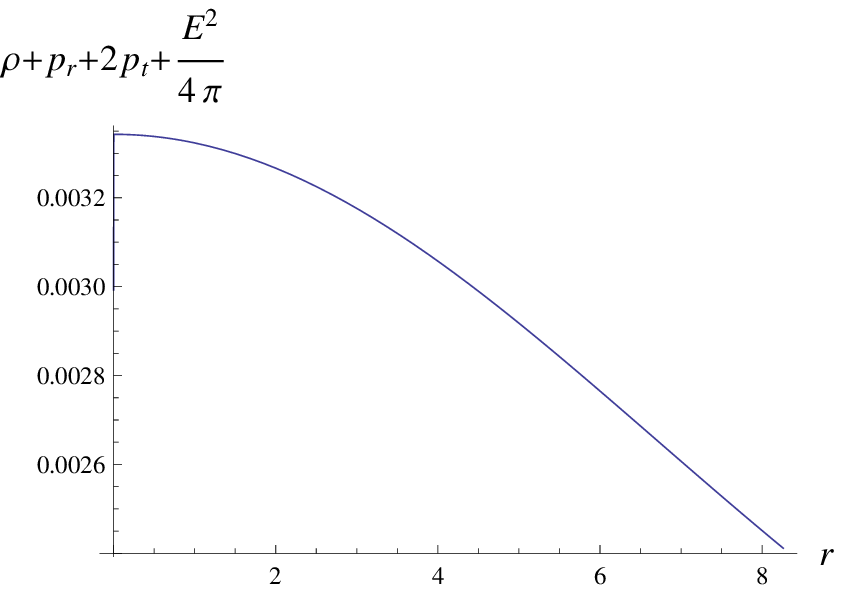, width=0.3\linewidth} \epsfig{file=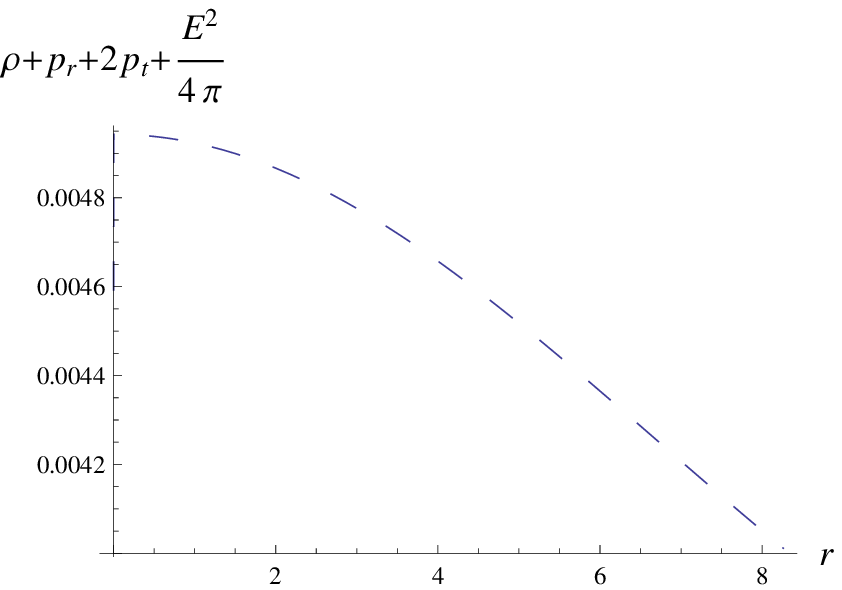,
width=0.3\linewidth}\caption{This figure represents the variation of
$\rho+p_r+2p_t+\frac{E^2}{4\pi}$ versus $r(km)$ for a strange star
of radius 8.26 km with $\beta=1$, $\beta=2$ and $\beta=3$.}
\end{figure}

\subsection{Energy conditions}

The anisotropic charged fluid considered as strange matter will
satisfy the null energy condition (NEC), weak energy condition (WEC)
and strong energy condition (SEC) if the following inequalities hold
simultaneously at all points for the given radius of the star:

\begin{equation}
\rho +\frac{E^{2}}{8\pi }\geq 0,
\end{equation}

\begin{equation}
\rho +p_{r}\geq 0,
\end{equation}

\begin{equation}
\rho +p_{t}+\frac{E^{2}}{4\pi }\geq 0,
\end{equation}

\begin{equation}
\rho +p_{r}+2p_{t}+\frac{E^{2}}{4\pi }\geq 0,
\end{equation}%
Employing these energy conditions at the centre $\left( r=0\right) ,$ we get
the following bounds on the constants $A$ and $B$:
\begin{equation*}
(i)NEC:\rho +\frac{E^{2}}{8\pi }\geq 0\Rightarrow A\geq 0.
\end{equation*}%
\begin{equation*}
(ii){WEC}:\rho +p_{r}\geq 0\Rightarrow A+B\geq 0,\rho +p_{t}+\frac{E^{2}}{%
4\pi }\geq 0\Rightarrow A+B\geq 0
\end{equation*}%
\begin{equation*}
(iii){SEC}:\rho +p_{r}+2p_{t}+\frac{E^{2}}{4\pi }\geq 0\Rightarrow
B\geq 0.
\end{equation*}%
Since $A>0$, so condition (i) is satisfied. The weak and strong
energy conditions (ii) and (iii) will be satisfied if $B\geq 0$.
With the set of values given in table \textbf{1} and \textbf{2}, we
have shown in Figures \textbf{6-9} that the energy conditions are
valid through out the interior of the given star.

\subsection{Stability}

In order to discuss the stability of our constructed models, we
follow the approach based on Herrera (1992) cracking (or
overturning) concept. According to this technique the squares of the
radial and tangential sound speeds must lies in the interval [0,1].
Further, cracking concepts predicts that the region for which the
radial speed of sound is greater than that of transverse speed ,
then such region is potentially stable region. It is clear that, for
no cracking (stability), the difference of two sound speeds,
i.e.,$v_{st}^{2}-v_{sr}^{2}$ should attain the same sign everywhere
inside the anisotropic matter distribution. In our model, we have

\begin{equation}
v_{sr}^{2}=\frac{dp_{r}}{d\rho }=\frac{1}{3},
\end{equation}

\begin{equation}
v_{st}^{2}=\frac{dp_{t}}{d\rho }=\frac{\alpha
+\gamma}{-3(A+B)rA\beta e^{-Ar^{2}}},
\end{equation}

where,

\begin{equation}
\alpha =e^{-Ar^{2}}\beta \left( 2B^{2}r-2ABr-\frac{2}{r^{3}}\right) +\frac{1%
}{r^{2}}
\end{equation}

\begin{equation}
\gamma =-2Are^{-Ar^{2}}\beta \left( \frac{7}{2}B-\frac{3}{2}%
A+B^{2}r^{2}-ABr^{2}+\frac{1}{r^{2}}\right)
\end{equation}

For causality condition to be satisfied, we must have

\begin{equation}
0<\frac{\alpha +\gamma }{-3(A+B)rA\beta e^{-Ar^{2}}}<1
\end{equation}

For the assumed set of values,  figure \textbf{10} shows that the
above condition is satisfied. Hence, our proposed model preserve
stability in $f(T)$ gravity .

\subsection{Surface Redshift }

We define the compactification factor as
\begin{equation*}
u=\frac{M_{eff}}{R}=\frac{1}{2}\beta \left( 1-e^{-Ar^{2}}\right)
\end{equation*}%
The surface red-shift $(Z_{s})$ corresponding to the above compactness $(u)$
is obtained as
\begin{equation*}
Z_{s}=\left(1-\beta \left(
1-e^{-Ar^{2}}\right)\right)^{-\frac{1}{2}}-1
\end{equation*}%
The maximum surface red-shift, in this set up, for a strange star of mass $%
1.4M_{\odot }$ and radius $8.26km$ turns out to be $Z_{s}=1.4$ for
$\beta=1$, while $Z_s<1.4$ for $\beta>1$ as shown in Figure
\textbf{11}.

\begin{figure}
\center\epsfig{file=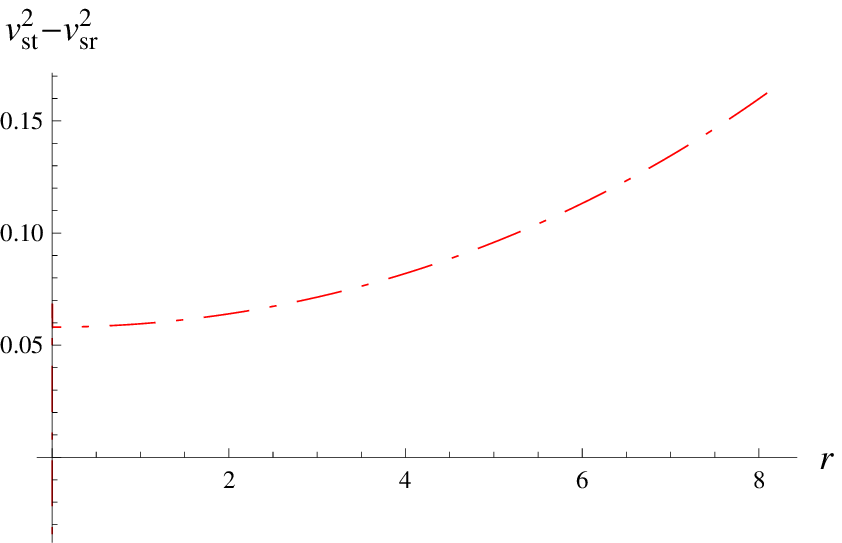, width=0.3\linewidth}
\epsfig{file=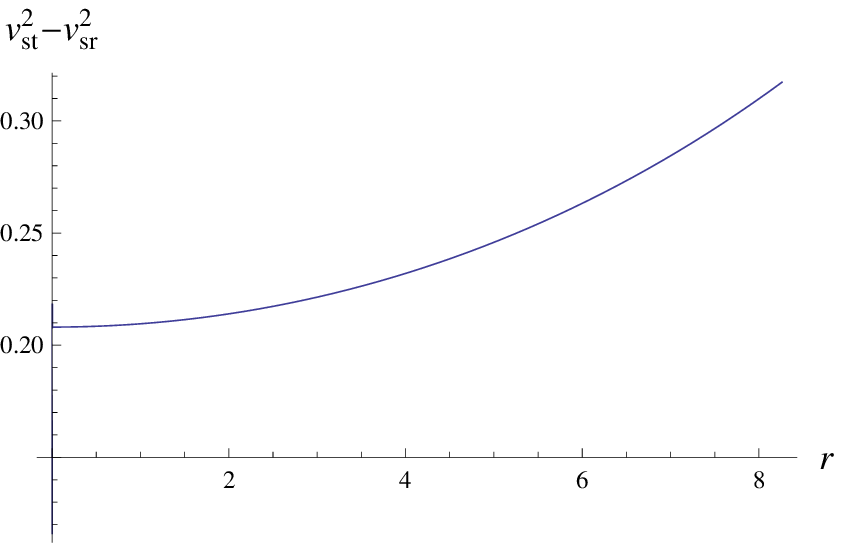, width=0.3\linewidth} \epsfig{file=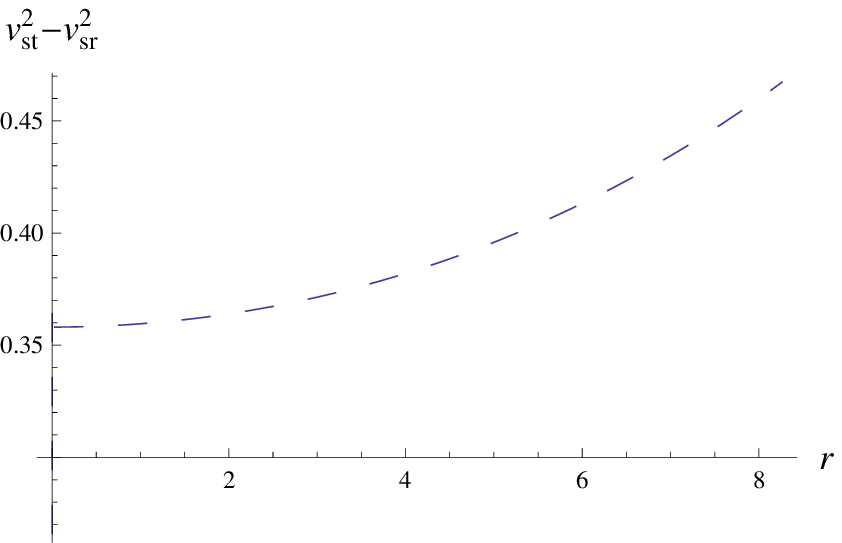,
width=0.3\linewidth}\caption{This figure represents the variation of
$v_{st}^{2}-v_{sr}^{2}$ versus $r(km)$ for a strange star of radius
8.26 km with $\beta=1$, $\beta=2$ and $\beta=3$.}
\end{figure}
\begin{figure}
\center\epsfig{file=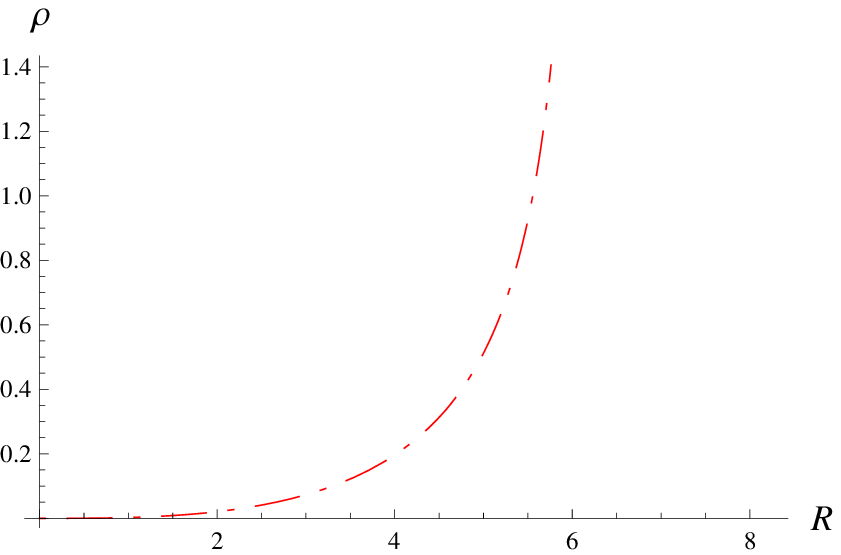, width=0.3\linewidth}
\epsfig{file=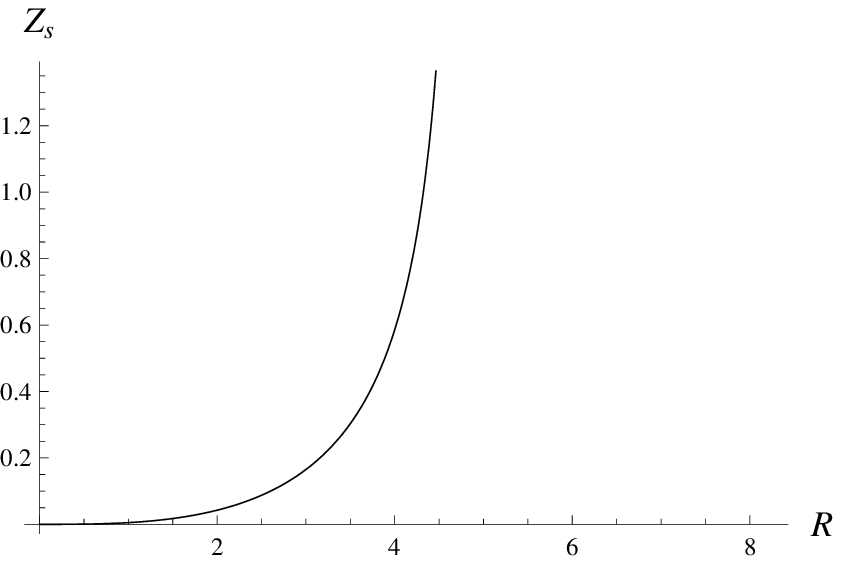, width=0.3\linewidth} \epsfig{file=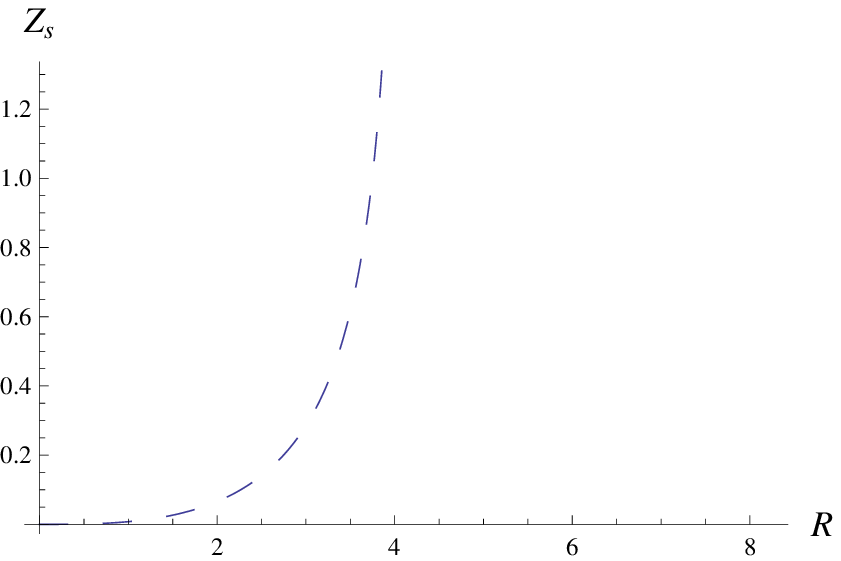,
width=0.3\linewidth}\caption{This figure represents the variation of
redshift $Z_s$ versus $r(km)$ for a strange star of radius 8.26 km
with $\beta=1$, $\beta=2$ and $\beta=3$.}
\end{figure}
\section{Discussion}
During the last few years, theoretical physicists have shown their
attraction to study gravitational field as the effect of torsion
instead of curvature of the underlying geometry. The torsion theory
was originally derived in parallel lines to general theory of
relativity which describes gravity in term of curvature. Hence, this
idea was promoted as teleparallel equivalence of GR. The black hole
(BH) and neutron stars are the highly dense astronomical objects
which store information about the entropy on the BH horizon. In the
modified TEGR, $f(T)$ gravity posses many significant feature as
compared to GR. Recent observations from solar system orbital
motions in order to constrain $f(T)$ gravity have been made and
interesting results have been found. The general constraints for the
existence/non-existence of compact stars have been studied in $f(T)$
gravity (Boehmer et al. 2011). It is interesting to drive the models
of the compact stars in $f(T)$ gravity using the diagonal tetrad
field with charged anisotropic fluid and MIT Bag model.

This paper deals with construction of analytic models of compact
stars in $f(T)$ gravity with charged anisotropic matter and MIT Bag
EoS. The inner region of the stars have been taken as static charged
anisotropic spherical source. By using the diagonal tetrad field,
the equations of motion have been formulated to complete the
discussion. One of the field equations implies that unknown function
$f(T)$ appears as a linear function of $T$ i.e., $f(T)=\beta T
+{\beta}_1$, where $\beta$ and ${\beta}_1$ are the constant of
integration. Using this form of $f(T)$ with KB metric functions, we
have determined explicitly the matter components and electric field
intensity. The anisotropy, regularity and energy conditions have
been discussed in detail. The observed values of masses, radii and
charge of compact stars have been used to calculate the unknown
constants of KB metric. The first and second derivatives of density
and pressures, implies the maximality of these quantities at the
center, and these have decreasing radial profile. The values of
density and pressure at $r=0$ and $r=R$ are given in tables
\textbf{3-5}.

We have found that (see Figure \textbf{4}) that there exist
repulsive gravitational force for $1\leq\beta\leq2$ as $(\Delta
>0)$, while for $\beta\geq3$ force becomes attractive. The first
case regarding to repulsive gravitational force leads to the
construction of more massive object, while second case gives the
formation of smaller objects. The Herrera's cracking concept states
that the region for which the radial speed of sound is greater than
that of transverse speed, then such region is potentially stable
region. It is clear that for stability, the difference of two sound
speeds, i.e.,$v_{st}^{2}-v_{sr}^{2}$ should attain the same sign
everywhere inside the anisotropic matter distribution. In our
discussion, the compact stars remains stable in $f(T)$ gravity even
in the presence of electromagnetic field. The maximum surface redshift, for a strange star of mass $%
1.4M_{\odot }$ and radius $8.26km$ in $f(T)$ gravity turns out to be
$Z_{s}=1.4$ for $\beta=1$, while $Z_s<1.4$ for $\beta>1$ as shown in
Figure \textbf{11}.

\section{Conflict of Interest}

The authors declare that there is no conflict of interest regarding
the publication of this work.\\

\end{document}